\documentclass[prb,twocolumn,showpacs,superscriptaddress,preprintnumbers,amssymb]{revtex4}
\usepackage{graphicx}
\usepackage{dcolumn}
\usepackage{bm}
\usepackage{wasysym}

\newcommand{\beq}{\begin{equation}}
\newcommand{\eeq}{\end{equation}}
\newcommand{\beqn}{\begin{eqnarray}}
\newcommand{\eeqn}{\end{eqnarray}}

\begin{document}

\title{Entanglement Entropy of Coupled Conformal Field Theories and Fermi Liquids}

\author{Cenke Xu}

\affiliation{Department of Physics, University of California,
Santa Barbara, CA 93106}

\date{\today}

\begin{abstract}

In this paper we calculate the entanglement R\'{e}nyi entropy of
two coupled gapless systems in general spatial dimension $d$. The
gapless systems can be either conformal field theories (CFT), or
Fermi liquids. We assume the two systems are coupled uniformly in
a $h-$dimensional submanifold of the space, with $0 \leq h \leq d
$. We will focus on the scaling of the R\'{e}nyi entropy with the
size of the system, and its scaling with the inter-system coupling
constant $g$. Three approaches will be used for our calculation:
({\it 1}) exact calculation with ground state wave-functional,
({\it 2}) perturbative calculation with functional path integral,
({\it 3}) scaling argument.

\end{abstract}
 \maketitle


\section{Define the Problem}

The entanglement entropy of a system $\mathcal{H}$, refers to the
entropy of the reduced density matrix of subsystem $\mathcal{A}$,
after tracing out the rest of the system
$\mathcal{H}/\mathcal{A}$. Usually $\mathcal{A}$ and
$\mathcal{H}/\mathcal{A}$ are simply separated spatially. If
$\mathcal{H}$ is a conformal field theory (CFT), it is
well-understood that for dimension $d > 1$, the leading
contribution to the entanglement entropy $S_\mathcal{A}$ is
proportional to the size of the boundary of subsystem
$\mathcal{A}$ \cite{srednicki1993}; while at $d = 1$,
$S_\mathcal{A}$ is proportional to $\ln(L_\mathcal{A})$,
$L_\mathcal{A}$ is the size of $\mathcal{A}$
\cite{cardy2004,wilczek1994,vidal2003}. If $\mathcal{H}$ is a
Fermi liquid, then the leading order entanglement entropy is
$S_\mathcal{A} \sim L_{\mathcal{A}}^{d - 1} \log (L_\mathcal{A})$
with arbitrary dimension $d$
\cite{wolf2006,klich2006,swinglefermi}.

In this work, we study the entanglement entropy between two
coupled systems $\mathcal{A}$ and $\mathcal{B}$ in general
dimension $d$, namely the entanglement entropy of the reduced
density matrix $\rho_\mathcal{A}$, after tracing out
$\mathcal{B}$. The entire action of the system can be
schematically written as \beqn \mathcal{S} = \int d\tau d^dx
\mathcal{L}_\mathcal{A} + \mathcal{L}_\mathcal{B} +
gO_\mathcal{A}(x)O_\mathcal{B}(x)V(x). \label{action}\eeqn
$O_\mathcal{A}$ and $O_\mathcal{B}$ are normal ordered operators
($\langle O_\mathcal{A} \rangle = \langle O_\mathcal{B} \rangle =
0$) of $\mathcal{A}$ and $\mathcal{B}$ respectively, $V(x)$ is a
function of space. The entanglement entropy is in general a
complicated function of $g$ (even for most of the exactly solvable
cases), so we will always assume a weak coupling $g$, and focus on
the scaling of the entropy with weak $g$. Our current work will
focus on the R\'{e}nyi entropy, which is defined as \beqn
S_{\mathcal{A}, n} = \frac{1}{1 - n} \log \left(
\mathrm{tr}\rho^n_\mathcal{A} \right). \eeqn We take $n = 2$ as an
example hereafter, but the results of our paper are insensitive to
$n$.

Our system is defined on a $d-$dimensional space with coordinates
$(x_1, \cdots, x_d)$, and $- L < x_i < L$. We assume $\mathcal{A}$
and $\mathcal{B}$ are coupled uniformly in a $h-$dimensional
submanifold $\mathcal{M}$ of the $d$ dimensional space, $i.e.$
\beqn && V(x_1, \cdots x_h, 0, \cdots, 0) = 1, \ \ \ V(x) = 0 \ \
\mathrm{otherwise}. \label{h}\eeqn When $h = d$, the coupling is
uniform in the entire space, while when $h = 0$ the coupling is
only through a point contact.

This paper is outlined as follows: In section II we will introduce
a general functional path integral and perturbation theory for the
entanglement entropy, and it will be applied to the coupled Fermi
liquids in section III. In section IV, we will study the
entanglement entropy of coupled CFTs. Besides the perturbation
theory, we will also introduce a scaling argument to understand
the qualitative behavior of the entanglement entropy. The results
with $h = d$ will also be checked with exact calculations based on
the ground state wave functionals of the coupled CFTs.

\section{Functional Path Integral and Perturbation Theory}

In most cases, the entanglement entropy between coupled systems
cannot be calculated exactly. In this section we will introduce a
general perturbation theory based on functional path integral
formalism for the entanglement entropy.

First of all, the reduced density matrix $\rho_\mathcal{A}$ of
subsystem $\mathcal{A}$ at zero temperature is \beqn \langle
\varphi_\mathcal{A} | \rho_{\mathcal{A}}|
\varphi^\prime_{\mathcal{A}} \rangle =
\mathrm{tr}_\mathcal{B}[\rho_\mathcal{AB}] \cr\cr = \lim_{\beta
\rightarrow \infty} \frac{1}{Z^{(\beta)}_{\mathcal{AB}}} \int
D\varphi_\mathcal{A} D\varphi_\mathcal{B}
|_{\varphi_\mathcal{A}(0) = \varphi_\mathcal{A}, \
\varphi_\mathcal{A}(\beta) = \varphi_\mathcal{A}^\prime} \cr\cr
\times \exp\left(- \int_0^\beta d\tau d^d x \
\mathcal{L}(\varphi_\mathcal{A}, \varphi_\mathcal{B})\right).
\eeqn $Z^{(\beta)}_{\mathcal{AB}}$ is the partition function of
the entire system: \beqn Z^{(\beta)}_{\mathcal{AB}} = \int
D\varphi_\mathcal{A} D\varphi_\mathcal{B} \exp\left(- \int_0^\beta
d\tau d^d x \ \mathcal{L}(\varphi_\mathcal{A},
\varphi_\mathcal{B})\right). \eeqn In the Lagrangian, if
$\varphi_\mathcal{A}$ and $\varphi_\mathcal{B}$ are boson
(fermion) fields, they are periodic (antiperiodic) in imaginary
time $\tau \in (0,\beta)$. In this section, $\varphi_\mathcal{A}$
and $\varphi_\mathcal{B}$ will be taken as boson fields for
example.

In order to calculate the R\'{e}nyi entropy with $n = 2$, we need
to evaluate the following quantity: \beqn
\mathrm{tr}[\rho^2_\mathcal{A}] = \lim_{\beta \rightarrow \infty}
\frac{1}{Z^{(\beta)2}_{\mathcal{AB}}} \int D\varphi_\mathcal{A}
D\varphi_\mathcal{B} \cr\cr \times \exp [ - \int d^d x \int_0^{2
\beta} d\tau \mathcal{L_A}(\varphi_\mathcal{A}) \cr\cr -
\left(\int_0^{\beta- \epsilon} + \int_{\beta + \epsilon}^{2\beta}
\right) d\tau \left( \mathcal{L_B}(\varphi_\mathcal{B}) + g
O_\mathcal{A} O_\mathcal{B} V(x)\right)]. \label{pi}\eeqn In the
numerator of this equation, we will keep \beqn &&
\varphi_\mathcal{A}(0) = \varphi_{\mathcal{A}}(2\beta), \cr\cr &&
\varphi_\mathcal{B}(0) = \varphi_{\mathcal{B}}(\beta - \epsilon),
\ \ \varphi_\mathcal{B}(\beta + \epsilon) =
\varphi_{\mathcal{B}}(2\beta ). \eeqn $\epsilon$ is an
infinitesimal positive number. Field $\varphi_\mathcal{B}$ is
integrated out in imaginary time segments $\tau \in (0, \beta -
\epsilon)$ and $\tau \in (\beta + \epsilon, 2\beta) $ separately.

The periodicity difference of $\varphi_\mathcal{A}(\tau)$ and
$\varphi_\mathcal{B}(\tau)$ is the key of this calculation.
Although we are always considering the case with zero temperature,
the limit $\beta \rightarrow + \infty$ should be taken {\it after}
all the calculations with finite $\beta$.

Now we try to calculate the entanglement entropy with perturbation
of $g$. Since we assumed that $O_\mathcal{A}$ and $O_\mathcal{B}$
are both normal ordered, the first order perturbation of $g$ of
Eq.~\ref{pi} vanishes. Expanding Eq.~\ref{pi} to the second order
of $g$, we obtain \beqn && \log \left( \mathrm{tr}
\rho^2_\mathcal{A}\right) \sim \lim_{\beta \rightarrow +\infty}
(S_1 - S_2), \cr\cr\cr S_1 &\sim& g^2 \left( \int_{0}^{\beta -
\epsilon} + \int_{\beta + \epsilon}^{2\beta} \right) d\tau_1
d\tau_2 d^dx_1d^d x_2 V(x_1) V(x_2) \cr\cr &\times&
G^{(2\beta)}_{\mathcal{AA}}(\tau_1, x_1, \ \tau_2, x_2)
G^{(\beta)}_{\mathcal{BB}}(\tau_1, x_1, \ \tau_2, x_2); \cr\cr\cr
S_2 &\sim& 2 g^2 \int_0^\beta d\tau_1 d\tau_2 d^dx_1 d^dx_2
V(x_1)V(x_2) \cr\cr &\times& G^{(\beta)}_{\mathcal{AA}}(\tau_1,
x_1, \ \tau_2, x_2) G^{(\beta)}_{\mathcal{BB}}(\tau_1, x_1, \
\tau_2, x_2). \label{perturbation} \eeqn $G_{\mathcal{AA}}$ and
$G_{\mathcal{BB}}$ are correlation functions: \beqn
G_{\mathcal{AA}} (\tau_1, x_1, \ \tau_2, x_2) &=& \langle
O_\mathcal{A}(\tau_1, x_1) O_\mathcal{A}(\tau_2, x_2) \rangle, \cr
\cr G_{\mathcal{BB}} (\tau_1, x_1, \ \tau_2, x_2) &=& \langle
O_\mathcal{B}(\tau_1, x_1) O_\mathcal{B}(\tau_2, x_2) \rangle,
\cr\cr G^{(2\beta)}(\tau_1, x_1, \ \tau_2, x_2) &=&
G^{(2\beta)}(\tau_1 + 2\beta, x_1, \ \tau_2, x_2), \cr\cr
G^{(\beta)}(\tau_1, x_1, \ \tau_2, x_2) &=& G^{(\beta)}(\tau_1 +
\beta, x_1, \ \tau_2, x_2). \eeqn Notice that there are two
different periodicities in these correlation functions. If
$O_\mathcal{A}$ and $O_\mathcal{B}$ are both bosonic operators,
then in the frequency space, $G^{(\beta)}$ has Matsubara frequency
$2\pi m/\beta$, while $G^{(2\beta)}$ has frequency $2\pi
m/(2\beta)$. For example, if $\mathcal{A}$ and $\mathcal{B}$ are
CFTs with $z = 1$, and we assume operator $O_\mathcal{A (B)}$ has
scaling dimension $\Delta_\mathcal{A (B)}$, then $G^{(\beta)}$
reads \beqn && G^{(\beta)}_{\mathcal{AA \ (BB)}}(0, 0, \ \tau, x)
\cr\cr &\sim& \frac{1}{\beta L^d} \sum_{\omega, k} \left(
\frac{1}{\omega^2 + k^2} \right)^{\frac{1}{2}(d + 1 - 2\Delta_{A
(B)})}e^{i\omega \tau + i\vec{k}\cdot \vec{x}}. \eeqn

Eq.~\ref{pi} and Eq.~\ref{perturbation} were formulated for
R\'{e}nyi entropy with $n = 2$ only, but their generalization to
arbitrary $n$ is straightforward.

\section{Entanglement entropy of coupled Fermi liquids}

\subsection{Uniform tunnelling}

In this section we will consider the entanglement entropy of
coupled Fermi liquids. The simplest situation that we can start
with, is that $\mathcal{A}$ and $\mathcal{B}$ are free Fermi gases
with $S^z = \pm 1/2$ respectively, and they are coupled together
through a uniform transverse magnetic field $HS^x$. Our goal is
calculate the entanglement entropy of $S^z = 1/2$ fermions, after
tracing out the $S^z = -1/2$ fermions. With uniform magnetic
field, this system can be trivially solved, and the reduced
density matrix $\rho_\mathcal{A}$ is a simple direct product of
the density matrix at each momentum $k$: \beqn \rho_\mathcal{A} =
\prod_k \otimes \rho_{\mathcal{A},k}. \eeqn If both $S^x = \pm
1/2$ spin states are occupied or unoccupied,
$\rho_{\mathcal{A},k}$ is a pure state density matrix. While if
only one of the spin states is occupied, $\rho_{\mathcal{A},k}$ is
maximally entangled: \beqn \rho_{\mathcal{A},k} = \frac{1}{2}
c^\dagger_{k,\uparrow}|0 \rangle \langle 0 | c_{k,\uparrow} +
\frac{1}{2} |0 \rangle \langle 0 |. \eeqn Therefore only the
states with energy $\varepsilon_f - H/2 < \varepsilon <
\varepsilon_f + H/2$ contribute to the entanglement entropy. Hence
the entanglement entropy should scale as \beqn S_\mathcal{A} \sim
\mathcal{N}(\varepsilon_f) |H| L^d. \label{fermiuniform}\eeqn
$N(\varepsilon_f)$ is the density of states at the Fermi surface.

\subsection{Point contact tunnelling}

Now suppose spin up and down fermions are coupled through a static
polarized magnetic impurity at $\vec{r} = 0$: $HS^x(0)$, this
impurity tunnels spin up and down fermions through the point
contact at $\vec{r} = 0$. The perturbation formalism developed in
the previous section is applicable here, as long as in the
calculation we keep the Matsubara frequency for spin up
($\mathcal{A}$) and down ($\mathcal{B}$) fermions as \beqn
\omega_\mathcal{A} = \frac{\pi(2m+1)}{2\beta}, \ \
\omega_\mathcal{B} = \frac{\pi(2n+1)}{\beta}. \eeqn Notice that
the difference between Matsubara frequencies $\omega_\mathcal{A}$
and $\omega_\mathcal{B}$ leads to \beqn \int_0^\beta d\tau
\exp\left(i(\omega_\mathcal{A} - \omega_\mathcal{B})\tau\right) =
\frac{i - (-1)^m}{\omega_\mathcal{A} - \omega_\mathcal{B}}, \eeqn
which contrasts the delta function in the usual case.

The leading order contribution to the entanglement entropy is a
straightforward application of Eq.~\ref{perturbation}, and it
leads to the following results: \beqn &&\log \left( \mathrm{tr}
\rho^2_\mathcal{A} \right) = \lim_{\beta \rightarrow
+\infty}(S^\prime_1 - S^\prime_2), \cr\cr\cr S^\prime_1 &\sim& H^2
\sum_{\omega_\mathcal{A}, \omega_\mathcal{B}}
\sum_{k_{\mathcal{A}}, k_{\mathcal{B}}} \frac{2 \ L^{-
2d}}{\beta^2 (\omega_\mathcal{A} - \omega_\mathcal{B})^2} \cr\cr
&\times& \frac{1}{ i \omega_\mathcal{A} - \varepsilon_{k,
\mathcal{A}} + \varepsilon_f } \frac{1}{ i \omega_\mathcal{B} -
\varepsilon_{k, \mathcal{B}} + \varepsilon_f }; \cr\cr\cr
S^\prime_2 &\sim&  2 H^2 \sum_{\omega} \sum_{k_{\mathcal{A}},
k_{\mathcal{B}}} L^{- 2d} \cr\cr &\times& \frac{1}{ i \omega -
\varepsilon_{k, \mathcal{A}} + \varepsilon_f } \frac{1}{ i \omega
- \varepsilon_{k, \mathcal{B}} + \varepsilon_f }.
\label{perturbative}\eeqn Frequency $\omega$ takes the usual
values $\pi (2m+1)/\beta$.

Correct evaluation of the frequency and momentum summation in
Eq.~\ref{perturbative} leads to the following result: \beqn
S_\mathcal{A} &\sim& H^2 \left( \int_{\varepsilon_f}^{+ \infty}
d\varepsilon_{\mathcal{A}} \int_{0}^{\varepsilon_f}
d\varepsilon_{\mathcal{B}} + \int_{0}^{\varepsilon_f}
d\varepsilon_{\mathcal{A}} \int_{\varepsilon_f}^{+ \infty}
d\varepsilon_{\mathcal{B}} \right) \cr\cr &\times&
\frac{1}{(\varepsilon_\mathcal{A} - \varepsilon_\mathcal{B})^2}
\mathcal{N}(\varepsilon_\mathcal{A})
\mathcal{N}(\varepsilon_\mathcal{B}).  \eeqn Since the density of
states is a constant close to the Fermi surface, this integral is
logarithmically divergent when $\varepsilon_{\mathcal{A}}$ and
$\varepsilon_{\mathcal{B}}$ are close to Fermi energy
$\varepsilon_f$. This logarithmic divergence will be cut-off by
$1/L$, thus the final result of the entanglement entropy is \beqn
S_\mathcal{A} \sim H^2 \left(\mathcal{N}(\varepsilon_f)\right)^2
\log(L). \label{fermiimpurity}\eeqn

In one dimension, the Fermi liquid becomes Luttinger liquid, which
is a CFT. Uniform magnetic field and point contact single fermion
tunnelling have scaling dimensions $\Delta = 1$ and $0$
respectively on a free fermion Luttinger liquid CFT, $i.e.$ the
point contact single fermion tunnelling is a marginal perturbation
on the free fermion Luttinger liquid. Later we will see that the
results in Eq.~\ref{fermiuniform} and Eq.~\ref{fermiimpurity} are
consistent with our general results about CFT with $d = 1$.

Many aspects of the Fermi liquid theory can be viewed as infinite
number of one dimensional fermions moving along the radial
direction, thus it is not surprising that the entanglement entropy
of Fermi liquid at higher dimension is qualitatively equivalent to
one dimensional free fermions. The connection between the Fermi
liquid and one dimensional CFT was also used to understand the
ordinary entanglement entropy of Fermi liquid \cite{swinglefermi}.

\subsection{$h-$dimensional tunnelling }

Now Let us assume the $S^z = \pm 1/2$ fermions are coupled through
a transverse magnetic field on a $h-$dimensional submanifold
$\mathcal{M}$ of the space (Eq.~\ref{h}). If we take the simplest
quadratic fermion dispersion, the second order perturbation in
Eq.~\ref{perturbation} gives the following result: \beqn
S_\mathcal{A} &=& S^\prime_1 - S^\prime_2, \cr\cr\cr S^\prime_1
&\sim& H^2 \sum_{\omega_\mathcal{A}, \omega_\mathcal{B}}
\sum_{k_i, k_{\mathcal{A}, j}, k_{\mathcal{B}, j}} \frac{2 \ L^{2h
- 2d}}{\beta^2 (\omega_\mathcal{A} - \omega_\mathcal{B})^2} \cr\cr
&\times& \frac{1}{i \omega_\mathcal{A} - \sum_{i = 1}^h k_i^2 -
\sum_{j = h + 1}^d k_{\mathcal{A} , j}^{2} + \varepsilon_f }
\cr\cr &\times& \frac{1}{i \omega_\mathcal{B} - \sum_{i = 1}^h
k_i^2 - \sum_{j = h + 1}^d k_{\mathcal{B} , j}^{2} +
\varepsilon_f}; \cr\cr\cr S^\prime_2 &\sim&  H^2 \sum_{\omega}
\sum_{k_i, k_{\mathcal{A}, j}, k_{\mathcal{B}, j}} L^{2h - 2d}
\cr\cr &\times& \frac{1}{i \omega - \sum_{i = 1}^h k_i^2 - \sum_{j
= h + 1}^d k_{\mathcal{A} , j}^{2} + \varepsilon_f} \cr\cr
&\times& \frac{1}{i \omega - \sum_{i = 1}^h k_i^2 - \sum_{j = h +
1}^d k_{\mathcal{B} , j}^{2} + \varepsilon_f}. \eeqn

When $h < d$, this integral is always logarithmically divergent,
thus the logarithmic contribution persists (at least to the second
order perturbation) as long as $h < d$: \beqn S_\mathcal{A} &\sim&
H^2 L^h \log(L).
\eeqn

\section{Entanglement entropy of coupled conformal field theories}

If the two coupled systems $\mathcal{A}$ and $\mathcal{B}$ are
both CFTs, the entanglement entropy due to coupling $g$ will
obviously depend on the scaling dimension $\Delta$ of the coupling
constant $g$. If the scaling dimensions of $O_\mathcal{A}$ and
$O_\mathcal{B}$ are $\Delta_\mathcal{A}$ and $\Delta_\mathcal{B}$
respectively, then the dimension of $g$ is $ \Delta = h + z -
\Delta_{\mathcal{A}}  - \Delta_{\mathcal{B}} $, $z$ is the
dynamical exponent.


\subsection{Exact calculation with ground state wave-functionals}

We will first consider the following theory \beqn \mathcal{L} &=&
\sum_k |\partial_\tau \varphi_{\mathcal{A},\vec{k}}|^2 +
|\partial_\tau \varphi_{\mathcal{B}, \vec{k}}|^2 + | k^z
\varphi_{\mathcal{A}, \vec{k}}|^2 + |k^z
\varphi_{\mathcal{B},\vec{k}}|^2 \cr\cr &+& g a \ k^m
|\varphi_{\mathcal{A},\vec{k}} + \varphi_{\mathcal{B},\vec{k}}|^2
+ g b \ k^m |\varphi_{\mathcal{A},\vec{k}} -
\varphi_{\mathcal{B},\vec{k}}|^2. \label{exact}\eeqn Both
$\mathcal{A}$ and $\mathcal{B}$ are free boson theories, and the
coupling between them is uniform in the $d-$dimensional space.
Here $a$ and $b$ are both dimensionless constants, and $g$ is a
small coupling constant. By adjusting number $m$, we can tune the
scaling dimension of $g$: $\Delta = 2z - m$. Since the entire
Lagrangian Eq.~\ref{exact} is quadratic, the entanglement entropy
$S_\mathcal{A}$ can be calculated exactly.

The entanglement entropy $S_\mathcal{A}$ can be calculated in the
same formalism as Ref.~\cite{kim2010}, where a marginal coupling
between two Luttinger liquids was considered ($z = 1 , \ m =2$).
Since the exact ground state wavefunctional of field
$\varphi_\mathcal{A}$ and $\varphi_\mathcal{B}$ can be written
down exactly, one can directly calculate the entropy with the
exact reduced density matrix. The coupling in Eq.~\ref{exact} is
uniform in space, it only couples $\varphi_\mathcal{A}$ and
$\varphi_\mathcal{B}$ modes with the same momentum, hence the
entropy is a simple sum of the entropy of coupled harmonic
oscillators at each momentum $k$. Since the exact result is in
general a complicated function of $g$, we will focus on the
leading term after expanding the exact result with small $g$.

The leading order results with different choices of $\Delta$ are
summarized as follows: \beqn (1), && \Delta = d/2, \cr\cr &&
S_{\mathcal{A}} \sim g^2 \log ( \frac{1}{|g|} ) L^d; \cr\cr (2),
&& \Delta
> d/2, \cr\cr  && S_{\mathcal{A}} \sim g^{d/\Delta} \ L^d,
\cr\cr (3), && \Delta < d/2, \cr\cr && S_{\mathcal{A}} \sim g^{2}
\ L^d. \label{exactresult}\eeqn Notice that $d/2$ is the critical
value of $\Delta$, when $\Delta < d/2$ the leading term of the
R\'{e}nyi entropy is always $g^2 L^d$, whether $g$ is relevant or
not. When $\Delta = d/2$, the leading order entropy acquires a
logarithmic correction. In Eq.~\ref{exactresult}, we have assumed
that the infrared cut-off of the system is $g$ instead of $1/L$,
$i.e.$ $L > g^{-1/\Delta}$. If $L < g^{-1/\Delta}$, the argument
of the logarithmic function in Eq.~\ref{exactresult} is replaced
by $L$.

It is known that the subleading correction to the boundary law of
the ordinary entanglement entropy contains important information
about the CFT \cite{ryuholo2,max2009}. In fact, in
Eq.~\ref{exactresult}, in addition to the leading term
proportional to the volume of the system, there are also
subleading terms. The subleading terms can be calculated
conveniently for free boson theory Eq.~\ref{exact} at $d = 1$, and
we summarize our results here: \beqn && S_{\mathrm{subleading}}
\sim \log\left( \frac{(\sqrt{a} + \sqrt{b})^4}{16ab} \right), \ \
\ \Delta
> 0, \cr\cr && S_{\mathrm{subleading}} \sim \Delta \log(L), \  \ \
b = 0, \ \ \Delta
> 0, \cr\cr && S_{\mathrm{subleading}} \sim g^2, \ \ \ \Delta = 0.
\label{subleading}\eeqn When $\Delta < 0$, there is no subleading
term at order $O(L^0)$.

Interestingly, when the coupling is relevant $i.e.$ $\Delta
> 0$, the most generic subleading contribution to the entropy approaches a constant
when $g \rightarrow 0$, as long as we take the
limit $L \rightarrow \infty$ first. If $b = 0$ ($i.e.$ the
coupling only affects mode $\varphi_\mathcal{A} +
\varphi_\mathcal{B}$, while $\varphi_\mathcal{A} -
\varphi_\mathcal{B}$ is still gapless), $S_{\mathrm{subleading}}$
is logarithmic of the system size, and its coefficient is a
universal constant proportional to the scaling dimension $\Delta$,
but it is independent of the magnitude of $g$.

The logarithmic subleading term in Eq.~\ref{subleading} may have
generalizations to other CFTs in one dimension. The universal
coefficient of the logarithmic term might be related to the
central charge of the CFT. We will study this general theory in
future.

\subsection{Perturbative calculation}

We can also apply our perturbative formalism
Eq.~\ref{perturbation} to the coupled CFTs. In the momentum and
frequency space, the entropy is evaluated as \beqn &&\log \left(
\mathrm{tr} \rho^2_\mathcal{A} \right) = \lim_{\beta \rightarrow
+\infty} (S^\prime_1 - S^\prime_2), \cr\cr\cr S^\prime_1 &\sim&
g^2 \sum_{\omega_\mathcal{A}, \omega_\mathcal{B}} \sum_{k_i,
k_{\mathcal{A}, j}, k_{\mathcal{B}, j}} \frac{4 \ L^{2h -
2d}}{\beta^2 (\omega_\mathcal{A} - \omega_\mathcal{B})^2} \cr\cr
&\times& \frac{1}{(\omega_\mathcal{A}^2 + \sum_{i = 1}^h k_i^2 +
\sum_{j = h + 1}^d k_{\mathcal{A} , j}^{2})^{\frac{1}{2}(d + 1 -
2\Delta_{\mathcal{A}})}} \cr\cr &\times&
\frac{1}{(\omega_\mathcal{B}^2 + \sum_{i = 1}^h k_i^2 + \sum_{j =
h + 1}^d k_{\mathcal{B},j}^{ 2})^{\frac{1}{2}(d + 1 -
2\Delta_{\mathcal{B}})}}; \cr\cr\cr S^\prime_2 &\sim&  g^2
\sum_{\omega} \sum_{k_i, k_{\mathcal{A}, j}, k_{\mathcal{B}, j}}
L^{2h - 2d} \cr\cr &\times& \frac{1}{(\omega^2 + \sum_{i = 1}^h
k_i^2 + \sum_{j = h + 1}^d k_{\mathcal{A} , j}^{2})^{\frac{1}{2}(d
+ 1 - 2\Delta_{\mathcal{A}})}} \cr\cr &\times& \frac{1}{(\omega^2
+ \sum_{i = 1}^h k_i^2 + \sum_{j = h + 1}^d k_{\mathcal{B},j}^{
2})^{\frac{1}{2}(d + 1 - 2\Delta_{\mathcal{B}})}}.
\label{perturbative2}\eeqn We have taken $z = 1$ as example.
$\Delta_{\mathcal{A (B)}}$ is the scaling dimension of
$O_{\mathcal{A (B)}}$. In this equation, \beqn \omega_\mathcal{A}
= \frac{2\pi (m + 1/2)}{\beta}, \ \ \ \ \omega, \
\omega_\mathcal{B} = \frac{2\pi n}{\beta}. \eeqn

Correct evaluation of the summation in Eq.~\ref{perturbative2}
will lead to the results consistent with the exact results
Eq.~\ref{exactresult}. In this calculation, one should always take
the limit $\beta \rightarrow \infty$ before the limit $L
\rightarrow \infty$. When $h = d$ (two CFTs are coupled uniformly
in the entire space), the leading contribution to
Eq.~\ref{perturbative2} is \beqn S_\mathcal{A} \sim g^2 L^d \int
\frac{1}{k^{2d + 2 - 2\Delta_\mathcal{A} - 2\Delta_\mathcal{B}}}
d^dk. \eeqn The scaling dimension of $g$ is $\Delta = d + 1 -
\Delta_\mathcal{A} - \Delta_\mathcal{B}$. If we take $\Delta =
d/2$, this integral gains a logarithmic contribution. Since
$\Delta > 0$, the higher order perturbation will acquire stronger
and stronger infrared divergence. If $L < g^{-1/\Delta}$, this
logarithmic divergence is cut-off by $1/L$; if $L >
g^{-1/\Delta}$, we expect the summation of the perturbation series
will eventually be cut-off by length scale $g^{-1/\Delta}$, 
so the final answer should be $S_\mathcal{A} \sim g^2 \log (1/|g|)
L^d$, which is consistent with Eq.~\ref{exactresult}.

Similarly, if we take $0 < h < d$, then when $\Delta = h/2$ the
perturbation theory gives the logarithmic term \beqn S_\mathcal{A}
\sim g^2 \log (\frac{1}{|g|}) L^h, \ \ (\mathrm{when} \ \Delta =
h/2). \eeqn

Now suppose we take $h = 0$ (point contact), the scaling dimension
of $g$ is $\Delta = 1 - \Delta_\mathcal{A} - \Delta_\mathcal{B}$.
For simplicity, we assume that $\Delta_{\mathcal{A}} =
\Delta_{\mathcal{B}} = \frac{1}{2}(1 - \Delta )$. Then
Eq.~\ref{perturbative2} is evaluated as 
\beqn S_\mathcal{A} &\sim& g^2 \int \frac{1}{k_1^{d + \Delta -
1}k_2^{d + \Delta - 1}} \frac{1}{(|k_1| + |k_2|)^2}d^dk_1 d^dk_2
\cr\cr &\sim& g^2 \int \frac{1}{k^{2\Delta + 1}} dk.
\label{marginalpoint}\eeqn When $\Delta < 0$, the point contact is
irrelevant, and this integral gives a constant result
$S_\mathcal{A} \sim g^2$, with a nonuniversal coefficient. When
$\Delta = 0$ (marginal point contact), we obtain a logarithmic
contribution, regardless of the total dimension $d$: \beqn
S_\mathcal{A} \sim g^2 \log(L). \label{marginalpoint2} \eeqn In
this case, since the point contact is marginal, the higher order
perturbation has the same logarithmic divergence as the leading
order, hence we expect this logarithmic divergence persists even
after we sum the entire series, and it can only be cut-off by $L$.

\subsection{Scaling Argument}

In this section we will try to understand the results obtained in
the previous two subsections using a simple scaling argument. The
argument in this section is a generalization of
Ref.~\cite{swingle2009}, where a scaling argument was introduced
to explain the logarithmic contribution of the ordinary
entanglement entropy of 1d CFT. Scaling argument was also
introduced to understand the entanglement entropy close to finite
temperature critical points \cite{hastings2011}.

We first note that, if the two coupled systems are gapped, the
R\'{e}nyi entropy is obviously proportional to the volume of the
coupled submanifold, and it should scale as $g^2$ for weak
coupling: $S_\mathcal{A} \sim g^2 L^h$. This is because
$S_{\mathcal{A}}$ is positive definite, hence the leading
contribution should be an even function of $g$. Also, the
entanglement entropy vanishes in the decoupling limit $g
\rightarrow 0$. Thus the leading analytic contribution from the
coupling should be $g^2$.

Now we consider two CFTs coupled on a $h-$dimensional submanifold
$\mathcal{M}$ of the space, and the entanglement entropy is
collected while coarse-graining the system. At length scale $l$,
the size of the coupled subsystem is effectively $L^h/l^h$. Within
length scale interval $d\left( \log l \right) $, the entanglement
entropy is \beqn d\left( S_\mathcal{A} \right) \sim
\frac{L^h}{l^h}g^2_l \ d\left(\log l \right). \eeqn $g_l$ is the
effective coupling constant at length scale $l$, and as long as
$g_l$ is weak, $g_l \sim g l^{\Delta}$. Thus if $\Delta > 0$, the
total entropy is \beqn S_\mathcal{A} \sim \int_{ l = a}^{l =
\mathrm{Min}[|g|^{-1/\Delta}, \ L]} d\left( \log l \right)
\frac{L^h}{l^h} g^2 l^{2\Delta}. \label{scaling}\eeqn This
integral gives us the following results for general $d \geq h >
0$, with different choices of $\Delta$: \beqn (1), && \Delta =
h/2, \cr\cr && S_{\mathcal{A}} \sim g^2 \log ( \frac{1}{|g|} )
L^h, \ \ \ L \gg g^{-1/\Delta}; \cr\cr && S_{\mathcal{A}} \sim g^2
L^h \log(L), \ \ \ L \ll g^{-1/\Delta}; \cr\cr (2), && \Delta >
h/2, \cr\cr  && S_{\mathcal{A}} \sim g^{h/\Delta} \ L^h, \ \ \ L
\gg g^{-1/\Delta}; \cr\cr && S_{\mathcal{A}} \sim g^2 L^{2\Delta},
\ \ \ L \ll g^{-1/\Delta}; \cr\cr (3), && \Delta < h/2, \cr\cr &&
S_{\mathcal{A}} \sim g^{2} \ L^h. \label{law}\eeqn Now the
critical value of $\Delta$ becomes $h/2$. When $h = d$, and $L
> g^{-1/\Delta}$, Eq.~\ref{law} reduces to the results
Eq.~\ref{exactresult} obtained from exact calculations.

In the special case with $h = 0$, $i.e.$ the CFTs are coupled
through a point contact, the integral in Eq.~\ref{scaling} gives
us the following results: \beqn \Delta = 0, && S_{\mathcal{A}}
\sim g^2 \log(L), \cr\cr \Delta > 0, && S_{\mathcal{A}} \sim C, \
\ \ L \gg g^{-1/\Delta}, \cr\cr && S_{\mathcal{A}} \sim g^2
L^{2\Delta}, \ \ \ L \ll g^{-1/\Delta}, \cr\cr \Delta < 0, &&
S_{\mathcal{A}} \sim g^2. \label{law4}\eeqn $C$ is a constant
which does not scale with $g$. When $\Delta = 0$, namely the case
with a marginal point contact, the leading contribution to the
entanglement entropy is a logarithmic term, which is independent
of the spatial dimension $d$. This conclusion confirms our
calculation in the previous subsection (Eq.~\ref{marginalpoint},
Eq.~\ref{marginalpoint2}), and confirms the calculation for Fermi
liquid with a point contact (Eq.~\ref{fermiimpurity}).

This simple scaling argument should be precise if $g$ is
irrelevant or marginal, for arbitrary $d$ and $h$. If $g$ is
relevant, the integral of length scale in Eq.~\ref{scaling} was
taken only from $a$ (lattice constant) to the scale where $g$
becomes nonperturbative. It seems like we have ignored the entropy
contribution {\it after} $g$ becomes nonperturbative. To
understand this problem, we need to know the long wavelength
properties of the system when $g$ is relevant, and there are two
possibilities:

{\it 1.} In most cases, a relevant coupling $g$ opens up a gap (or
local gap) for CFTs $\mathcal{A}$ and $\mathcal{B}$, namely all
the correlation functions $G(\tau_1 - \tau_2, x_1 - x_2)$ with
$x_1, x_2 \in \mathcal{M}$ decays exponentially when $|\tau_1 -
\tau_2| \rightarrow \infty$. In this case, the system can be
driven into either a direct product state between $\mathcal{A}$
and $\mathcal{B}$, or a maximally or partially entangled state
between $\mathcal{A}$ and $\mathcal{B}$ with a saturated
entanglement. For instance, if $\mathcal{A}$ and $\mathcal{B}$ are
free boson fields, and in the submanifold $\mathcal{M}$ they are
coupled as $- ag(\varphi_\mathcal{A} + \varphi_{\mathcal{B}})^2 -
bg(\varphi_\mathcal{A} - \varphi_{\mathcal{B}})^2$, then as long
as $ag
> 0$ and $bg > 0$, the relevant coupling $g$ will keep $\varphi_\mathcal{A}
= \varphi_\mathcal{B} = 0$ in $\mathcal{M}$, hence the system
becomes a direct product state between $\mathcal{A}$ and
$\mathcal{B}$ in the long wavelength limit. Then the entropy with
length scale $l > g^{-1/\Delta}$ is ignorable, and there is no
correction to Eq.~\ref{scaling}.

When $g$ drives the system into a maximally or partially entangled
state, then we need to add another contribution to the entropy,
which is \beqn S^\prime_\mathcal{A} \sim
\frac{L^h}{(g^{-1/\Delta})^h} = g^{h/\Delta} L^h. \eeqn With this
extra contribution, our results in Eq.~\ref{law} and
Eq.~\ref{law4} still hold.

{\it 2.} If some of the correlation functions in the coupled
submanifold $\mathcal{M}$ remain power-law even with relevant $g$,
then there is a residual scaling invariance after $g$ becomes
nonperturbative. In this case, we need to include the following
extra contribution to Eq.~\ref{scaling}: \beqn
S^\prime_\mathcal{A} \sim \int_{ l = g^{-1/\Delta}}^{l = L}
d\left( \log l \right) \frac{L^h}{l^h}. \eeqn This integral does
not modify any of the leading order terms in Eq.~\ref{law} with $h
> 0$, but it leads to a logarithmic contribution to Eq.~\ref{law4}
with $\Delta > 0$ and $h = 0$, $i.e.$ it only affects the case
with a relevant point contact coupling between $\mathcal{A}$ and
$\mathcal{B}$. For instance, in Eq.~\ref{exact}, although the mode
$\varphi_\mathcal{A} - \varphi_\mathcal{B}$ remains gapless when
$b = 0$, the exact results Eq.~\ref{exactresult} always agree with
Eq.~\ref{law} obtained from scaling integral Eq.~\ref{scaling} for
$h = d$, no matter $b = 0$ or not.

\section{Summaries and Extensions }

In this work, we studied the entanglement entropy of coupled Fermi
liquids and CFTs. Three different methods were used for the
calculation: perturbation theory, scaling argument, and exact
ground state wave-functional. These three approaches are
consistent with each other for all the cases that we can check.

It has been demonstrated that the holographic method is a very
powerful way of calculating the entanglement entropy
\cite{ryuholo1,ryuholo2} of CFT, assuming there is a bulk AdS
space duality of the boundary CFT. The ordinary entanglement
entropy is related to the area of the minimal surface of the bulk
AdS space. In future, we will try to develop a holographic
formalism to produce the results in the current paper. Since
AdS/CFT duality effectively ``geometrizes" the RG flow at the
boundary CFT theory, we expect the holographic calculation of the
entanglement entropy to be qualitatively equivalent to the scaling
argument discussed in this paper.

Significant progresses have been made in numerical simulation of
quantum many-body states. For instance, the multi-scale
entanglement renormalization ansatz (MERA) is especially powerful
in simulating one dimensional CFT \cite{vidal1,vidal2}. In future,
it will also be interesting to verify the conclusions in our paper
numerically.

\begin{acknowledgements}

The author thanks Andreas W. W. Ludwig, Guifre Vidal, Yong-Baek
Kim, and Matthew B. Hastings for very helpful discussions.

\end{acknowledgements}

\bibliography{entropy}

\end{document}